\documentclass[
twocolumn,
]{ceurart}

\usepackage{listings}
\usepackage{framed,multirow}

\usepackage{mathrsfs}
\usepackage{amssymb}
\usepackage{amsmath}
\usepackage{amsthm}
\usepackage{latexsym}

\usepackage{url}
\usepackage{xcolor}
\usepackage{multirow}
\usepackage{caption}
\usepackage{subcaption}
\usepackage{adjustbox}
\usepackage{mathrsfs}
\usepackage{xcolor}
\usepackage{hyperref}
\usepackage{graphicx}       

\usepackage{algorithm}
\usepackage{algorithmic}
\usepackage{tabularx,ragged2e}

\DeclareMathOperator*{\argmax}{arg\,max}

\usepackage{microtype}
\usepackage{mdframed}
\mdfdefinestyle{MyFrame}{%
    linecolor=gray,
    innertopmargin=2pt,
    innerbottommargin=2pt,
    innerrightmargin=2pt,
    roundcorner=10pt,
    innerleftmargin=2pt,
    linewidth=1pt,
    backgroundcolor=white
    }
    
\newtheorem{definition}{Definition}

\definecolor{codegreen}{rgb}{0,0.6,0}
\definecolor{codegray}{rgb}{0.5,0.5,0.5}
\definecolor{codepurple}{rgb}{0.58,0,0.82}
\definecolor{backcolour}{rgb}{0.95,0.95,0.92}
\definecolor{lightgray}{gray}{0.9}   

\definecolor{yellow}{RGB}{255,255,153}
\definecolor{grey}{RGB}{224,224,224}

\newboolean{showcomments}
\setboolean{showcomments}{true}
\ifthenelse{\boolean{showcomments}}
 { \newcommand{\mynote}[2]{
      \fbox{\bfseries\sffamily\scriptsize#1}
        {\small$\blacktriangleright$\textsf{\emph{#2}}$\blacktriangleleft$}}}
        { \newcommand{\mynote}[2]{}}

\definecolor{DarkOrange}{rgb}{0.8,0.3,0.0} 
\definecolor{DarkCyan}{rgb}{0.0, 0.55, 0.55}
\definecolor{Gray}{gray}{0.9}

\begin{document}

\copyrightyear{2022}
\copyrightclause{Copyright for this paper by its authors.
  Use permitted under Creative Commons License Attribution 4.0
  International (CC BY 4.0).}

\conference{The Thirty-Seventh AAAI Conference on Artificial Intelligence (AAAI-23) - SafeAI Workshop}

\title{On Evaluating Adversarial Robustness of Chest X-ray Classification: Pitfalls and Best Practices}



\author[1]{Salah GHAMIZI}[%
orcid=0000-0002-0738-8250,
email=salah.ghamizi@uni.lu,
]
\cormark[1]

\author[1]{Maxime CORDY}[%
orcid=0000-0001-8312-1358,
email=maxime.cordy@uni.lu,
]
\author[1]{Mike PAPADAKIS}[%
orcid=0000-0003-1852-2547,
email=michail.papadakis@uni.lu,
]
\author[1]{Yves LE TRAON}[%
orcid=0000-0002-1045-4861,
email=yves.letraon@uni.lu,
]

\address[1]{University of Luxembourg}

\cortext[1]{Corresponding author.}

\begin{abstract}
  Vulnerability to adversarial attacks is a well-known weakness of Deep Neural Networks. While most of the studies focus on natural images with standardized benchmarks like ImageNet and CIFAR, little research has considered real world applications, in particular in the medical domain.

    Our research shows that, contrary to previous claims, robustness of chest x-ray classification is much harder to evaluate and leads to very different assessments based on the dataset, the architecture and robustness metric. We argue that previous studies did not take into account the peculiarity of medical diagnosis, like the co-occurrence of diseases, the disagreement of labellers (domain experts), the threat model of the attacks and the risk implications for each successful attack.      
    
    In this paper, we discuss the methodological foundations, review the pitfalls and best practices, and suggest new methodological considerations for evaluating the robustness of chest xray classification models. Our evaluation on 3 datasets, 7 models, and 18 diseases is the largest evaluation of robustness of chest x-ray classification models. 
\end{abstract}


\maketitle

\section{Introduction}

Chest x-ray (CXR) is an affordable, easy-to-use medical imaging and diagnostic technique. Chest radiography is the most requested radiologic examination. It is commonly used to diagnose a broad range of lung diseases and abnormalities, such as Atelectasis, Pneumothorax, and even early lung cancer. The chest film reading consists of identifying areas of increased density or areas of decreased density. The areas are identified with different shades of grey on the grayscale images. 
Practitioners commonly use one or two views in CXR. The postero-anterior (PA) view is the front view. Examining all areas where the lung borders the diaphragm, the heart, and other mediastinal structures is essential. The lateral view, called the anteroposterior (AP) view, can be used in addition to refining the diagnosis.

Although disease patterns may seem well-defined, correctly interpreting the CRX films is always a significant challenge, even for radiologists. Families overlap and sometimes are concurrent. In addition, imaging processing provides various grades of contrast levels and is not exempt from noise. Therefore, examining one CXR film can be misleading and may even cause diagnostic discrepancies from one practitioner to another.    
Medical errors, especially diagnostic errors, account for additional medical spending of \$17 to \$29 billion \cite{kohn2000errors}. Garland \cite{garland1959studies} reported a 32\% retrospective error rate in the interpretation of abnormal CXR, while the daily error rate averaged only 3\% to 4\% when negative studies were included. More recent studies have shown that misdiagnosis errors of chest x-ray images remain high even with the advances in practice and imaging systems \cite{biais}.

The challenge of providing a reliable and efficient diagnosis has motivated increasing research for automated diagnosis systems. While the first attempt for an automated CXR diagnosis system started in the 1960s \cite{firstcxr}, recent techniques using Deep Learning have shown promising performance \cite{rajpurkar2017chexnet,yao2019strong}. Riverain and Delft imaging systems have already developed many commercial products \cite{Qin2018ComputeraidedDI}, and some have even obtained FDA clearance for large-scale commercialization, such as Zebra Medical Vision. 

While these systems provide remarkable figures in their respective studies, recent research has shown generalization issues \cite{yao2019strong, pooch2020trust,baltruschat2019comparison}. Some have proposed a few hypotheses to explain the discrepancies: Errors in labeling \cite{oakdenrayner2019exploring}, practitioner biases and disagreements \cite{biais}, and more generally, overfitting of models and lack of generalization across multiple datasets \cite{cohen2020limits}.

A new facet of deep learning generalization has emerged in recent years. The so-called "adversarial examples" have exposed the inherent vulnerability of machine learning models in general and deep learning image classification models in particular to small perturbations. Especially inputs that have been engineered to cause misclassification. The study of the adversarial vulnerability of image classification models has only recently tackled medical systems. However, the few studies of chest x-ray classification robustness \cite{advattackmedicalsystem, taghanaki2018vulnerability,understandingadvattacks,robustdetection} has focused on binary classification (normal VS disease) and drew conclusions from one dataset and one or two models. 

However, we argue that natural image classification setting and the medical classification setting are very different and require the evaluation of different threat models, robustness metrics, and hyper-parameters.  

To uncover the inconsistencies between the two settings, we provide the first large-scale study of chest x-ray classification vulnerability to the best of our knowledge. 
Furthermore, we introduce two novel methodological considerations for evaluating robustness in medical domains: cross-domain generalization and domain-specific knowledge. We argue that a rigorous evaluation of the robustness of medical classifiers in general and chest x-ray classifiers in particular needs to consider these facets.


To summarize, our contributions are:
\begin{itemize}
    \item We survey the literature on adversarial robustness in chest x-ray classification and identify the major pitfalls and limitations. 
    \item We propose a set of principles and recommendations for how such pitfalls could be mitigated. 
    \item We demonstrate the impact and criticality of the principles through an empirical study of chest x-ray classification robustness using three datasets, seven models, and 18 diseases.
\end{itemize}



\section{Related Work}

\paragraph{Adversarial attacks}

An adversarial attack is the process of intentionally introducing perturbations to the inputs of a machine learning model to cause wrong predictions. One family of adversarial attacks is \emph{poisoning attacks}~\cite{biggio2012poisoning} where the inputs targeted are the training set and occur during the learning step, while \emph{evasion attacks} \cite{Biggio2013} focus on the inference step. 

One of the earliest attacks is the Fast Gradient Sign Method (FGSM) \cite{fgsm_goodfellow2015explaining}. It adds a small perturbation \(\eta\) to the input of a neural network, which is defined as:

\begin{equation}
\eta = \epsilon \, \text{sign}(\nabla_x \mathscr{L}_i(\theta,x,y_i)),\label{eq:fgsm}
\end{equation}
where \(\theta\) are the parameters of the network, \(x\) is the input data, \(y_i\) is its associated target, \(\mathscr{L_i}(\theta,x,y_i)\) is the loss function used, and \(\epsilon\) the strength of the attack. 
Following Goodfellow, other attacks were proposed, first by adding iterations~\cite{kurakin2016adversarial}, projections and random restart~\cite{pgd_madry2019deep}, momentum~\cite{dong2018boosting}, adaptive steps~\cite{autoattack} and constraints~\cite{simonetto2021unified}.

Recent work investigated attacks for finance~\cite{coeva2}, privacy~\cite{Ghamizi_2021_ICCV}, and navigation~\cite{Ghamizi_Cordy_Papadakis_Traon_2022}, and demonstrated that real-world attacks require special considerations.

\paragraph{Adversarial attacks for CXR disease classification.} 

Taghanaki et al. \cite{asgari2018vulnerability} were among the first to evaluate the robustness of CXR image classification against adversarial examples. They evaluated white box and black box attacks on two binary neural networks (ResnetV2 and NasNet Large) using the ChestX-ray14 dataset \cite{Wang_2017}. They showed that both models are vulnerable against gradient-based attacks (100\% success rate of attacks). While their evaluation pioneered the research on adversarial attacks in the medical setting, their evaluation focused on binary classification in a restricted setting. 

Finlayson et al. \cite{finlayson2018adversarial} focused on binary image classification for medical diagnosis. Their study covered CXR, Fundoscopy, and Dermoscopy diagnosis, and they also showed that PGD attacks achieved a 100\% success rate on the ChestX-ray14 Pneumothorax label using one model.

Ma et al. \cite{ma2021understanding} had another take on the robustness of CXR image classification models. They compared the robustness of binary classification, 3-label, and 4-label classification. They showed that while PGD had a success rate of over 99\% on all of them, the vulnerability seems to decrease with the increased number of labels. The three classifiers were trained on the ChestX-ray14 dataset, each time with a subset of labels. Our study covers the complete scenario of 18-label classifiers trained on different datasets (and distributions) and architectures. A previous study \cite{cohen2020limits} showed essential performance differences between the different labels that can explain the slight variation of robustness across the set of labels. Some labels are already challenging to learn and, similarly, to attack. Our evaluation, on the contrary, shows that the variations between different datasets and architectures are significant and that some models are actually resilient against adversarial attacks.
\section{Pitfalls and principles of chest x-ray robustness evaluations}

\subsection{Medical images differ from natural images}

Before we evaluate common practices, it is insightful to understand why using the experimental protocol of adversarial attacks on natural image datasets is rarely relevant in the context of chest X-ray classification.   

The first consideration is the nature of the tasks and labels. In ImageNet\cite{deng2009imagenet} and Cifar\cite{cifar10} classification, the images are designed to highlight one class (the ground truth class) more than the others. Meanwhile, chest radiographs are real images in which the same image can contain multiple diseases of equal importance. Chest X-ray classification can be seen as a multi-label classification problem, and using metrics and losses specific to this field of machine learning can provide a more faithful representation of the robustness of the models.

Another consideration is that the labels of the images and their probabilities are subjective to the radiologists who provided the ground truths. Cohen et al. \cite{biais} have shown that radiologists suffer from an \emph{availability bias}: They judge the probability of an event by the ease with which examples come to one's mind. In addition, radiologists also exhibit a \emph{confirmation bias}. They actively search for data to confirm a specific hypothesis rather than looking for data that facilitate efficient testing of a competing hypothesis \cite{biais}. Furthermore, Cohen et al.\cite{cohen2020limits} have shown that there is a large discrepancy in the agreement on the most probable diseases across different datasets (and thus the labelers). Testing the robustness of the model when the actual ground truth is uncertain is an arduous task. A chest radiograph image that can be considered adversarial by a practitioner can be considered legitimate by another. We can mitigate the risk of consistency by considering the top-k predicted labels and ensuring that they match the consensus among the practitioners. This consideration requires thus new definitions of adversarial examples in the medical setting.

Additionally, the risk associated with an error has a different impact depending on the nature of the error. There are two risks in medical diagnosis: \emph{misses}, i.e., when the classifier does not detect the correct disease among the most probable classes, and \emph{misinterpretations} when the most probable disease leads to an incorrect diagnosis. The latter can have a different impact depending on how similar the predicted labels from the original ones. Similarity can take into account the treatment process: Confusing 2 diseases that, in the end, require similar treatment is less detrimental than confusing two diseases with different treatments. Similarity can also be considered following disease taxonomy: Diseases that belong to the same families/branches can be considered more similar. The four pattern approach commonly used \cite{DELACEY20082,diff_diagnosis} considers four families: Consolidation, Interstitial, Nodules or masses, and Atelectasis. Within each family, there is a wide range of diseases. Confusing a disease from one family with one from another can be prejudicial. Some diseases regularly occur together \cite{rajpurkar2017chexnet} and thus can be used to diagnose each other. A misclassification that confuses them is less prejudicial than confusing two improbable diseases.

\subsection{Literature review}

While previous studies \cite{apostolidis2021survey, kaviani2022adversarial} referenced the major publications about adversarial robustness in the medical setting, their work was an index of the literature and not a critical analysis of the protocol or the relevance and impact of the experimental designs. 

\paragraph{Collection protocol.} Starting from the two existing surveys, we collected the publications that have been peer-reviewed related to CXR classification from 2018. There are, in total, 16 publications that match this scope. For each publication, we record seven criteria that, when not sufficiently evaluated, can lead to overestimated or even wrong claims. We summarize this literature in Table \ref{table:rq0-practices}. We detail each of the criteria below.

\begin{table*}[]
\scalebox{0.8}{
\begin{tabularx}{1.25\linewidth}{l|X|X|X|X|X|X}
Reference   & Datasets    & Threat models & Architectures     & Robust models    & Attacks    & Metrics
\\
\hline
Finlayson et al. \cite{finlayson2018adversarial}  & Binary NIH  & Whitebox, Graybox & Resnet50   & No& PGD, Patch& Accuracy, AUC   
\\\hline
Ma et al. \cite{ma2021understanding}   & 4 class NIH & Whitebox& Resnet50   & No& FGSM, CW, BIM, PGD   & Accuracy, AUC   
\\\hline
Yao et al. \cite{yao2019strong} & Binary Pneumonia & Whitebox& Resnet50, VGG-16& No& FGSM, B/MIM, PGD    & Accuracy    
\\\hline
Tian et al. \cite{tian2021bias} & Binary Pneumonia & Whitebox, Graybox& Resnet, DenseNet, MobileNet & No& FGSM, CW, PGD, B/MIM, Custom  & Success Rate 
\\\hline
Hirano et al. \cite{hirano2020vulnerability}    & 3 class COVID    & Whitebox& CovidNet   & Adversarial  Retraining  & FGSM, PGD  & Accuracy    
\\\hline
Pal et al. \cite{pal2021vulnerability}  & Binary COVID&  Whitebox  & VGG16, InceptionV3   & No& FGSM  & Accuracy    
\\\hline
Gongye et al. \cite{gongye2020new}   & 3 class COVID    & Whitebox& Resnet18   & No& FGSM, PGD2 & Accuracy    
\\\hline
Rahman et al.  \cite{rahman2020adversarial}  & Binary COVID& Whitebox, Blackbox API & Resnet50   & No& FGSM, PGD, DeepFool, +4 & Loss   
\\\hline
Taghanaki et al. \cite{taghanaki2018vulnerability} & Binary NIH  & Whitebox, Graybox & Inception, NasNet-Large& No& FGSM, PGD, DeepFool, + 6  & Accuracy, AUC   
\\\hline
Anand et al.  \cite{anand2020self}& Binary Pneumonia & Whitebox& VGG11 & Adversarial  Training    & FGSM, PGD  & AUC   
\\\hline
Kovalev et al.  \cite{kovalev2019influence}   & Binary Custom    & Whitebox& InceptionV3& No& PGD   & Success rate 
\\\hline
Hirano et al.  \cite{hirano2021universal}    & Binary Pneumonia & Whitebox, Graybox & ResNet, VGG, DenseNet & Adversarial  Retraining  & FGSM, DeepFool  & Success rate, confusion matrix 
\\\hline
Xue et al.  \cite{xue2019improving} & 3 class RSNA& Whitebox& ResNet18, VGG16 & Custom denoiser  & FGSM, BIM,  CW  & Accuracy    
\\\hline
Tripathi et al  \cite{tripathi2020fuzzy}  & 3 class COVID    & Whitebox& ResNet18, VGG16 & FUIT Adversarial train   & FGSM, BIM,  CW, PGD  & Accuracy    
\\\hline
Xu et al  \cite{xu2021towards}   & NIH    & Whitebox& DenseNet-121    & Adv training  & PGD, GAP  & Success Rate, AUC, Accuracy   
\\\hline
Li et al  \cite{li2020robust}   & NIH    & Whitebox& DenseNet-121    & Detection   & FGSM, BIM, PGD  & N/A    
\\ \hline
\end{tabularx}
}
\caption{Peer-reviewed publications about adversarial robustness in CXR classification from 2018 to 2021
}
\label{table:rq0-practices}
\vspace{-4mm}
\end{table*}

\paragraph{Datasets.} The selection of datasets entails two hazards that can affect the conclusions. First, the evaluation of binary classification (9 publications among the 16) leads to an overestimation of the robustness of the models. Indeed, attacking a multi-label classifier is much easier \cite{madry2017towards} as the decision boundaries are more blended than single-label classifications. Another risk arises when drawing conclusions about CXR classification from one dataset only. All publications we identified restrict their evaluation to \textbf{one} CXR dataset. We demonstrate empirically that the conclusions about the robustness of a model significantly differ from one CXR dataset to another.

\paragraph{Threat models.} 
The evaluation of the whitebox setting is relevant to understand the internals of the DNN model or to evaluate the worst-case scenario. However, in practice, access to the model and dataset of a specific hospital/practitioner is unrealistic—only five papers evaluated a more realistic setting, with at least the graybox attack scenario. 
Our results demonstrate that the conclusions can change when assessing realistic cases where the attacker only has access to the target dataset (graybox) or even no knowledge (blackbox).

\paragraph{Architectures.} 
Nine papers among 16 restricted the robustness evaluation to only one CXR architecture. We demonstrate that the robustness of architectures can significantly vary with the threat model and the dataset under evaluation.

\paragraph{Robust models.}
This criterion is critical, as demonstrated by Carlini et al. in multiple publications \cite{athalye2018obfuscated,he2017adversarial,carlini2019evaluating}. Since 2018, robustification solid protocols have been designed using adversarial training, and multiple repositories of robust models are available (Robustbench, for example, \cite{croce2020robustbench}). Unfortunately, only two publications (\cite{anand2020self, xu2021towards} considered strong defenses, and five others used broken or weak defenses.

\paragraph{Strong attacks.}
Fourteen publications investigated potentially strong attacks (CW, PGD), and their evaluation used very few iterations and a limited perturbation budget. Although current good practices are to use adaptive and robust attacks such as AutoAttack \cite{autoattack}, we show empirically that increasing PGD budgets already leads to surprising behaviors when comparing datasets and architectures.

\paragraph{Evaluation attacks.}
We demonstrate that the success rate and accuracy of adversarial examples are misleading because of the nature of CXR classification. For example, the co-occurrence of pathologies and the risk associated with each type of error lead to alternative conclusions in the evaluation. We propose a new RISK metric to take into account the specificity of CXR classification.

\section{Empirical evaluation}
\label{sec:robustness-metrics}

In traditional adversarial attack literature, we evaluate the robustness using the success rate of the attacks, i.e. 1-accuracy of the predictions over the adversarial examples (generally called \emph{robust accuracy}. The success rate and the robust accuracy has directly been used in previous literature about CXR adversarial examples \cite{paschali2018generalizability, robustdetection, understandingadvattacks, taghanaki2018vulnerability}. We argue that the specificities of medical classification in general, and CXR image classification in particular, make these metrics irrelevant. First, some datasets are provided as multi-label datasets (NIH for instance) and multiple diseases can occur together. Other datasets are built around the uncertainty of diagnosis, when the domain experts do not provide the same diagnosis for a given input. CheXpert dataset, for instance, has been designed with 3 values of labels: positive (1), uncertain (-1) and negative (0). Finally, \cite{cohen2020limits} have showed that 2 models trained for the same task on a different dataset have different degrees of agreement of the most probable labels and diagnosis.
To take into account the uncertainty and co-occurrences of labels, we propose to use a \textbf{k-robust accuracy}.

\begin{definition}

Let $\mathscr{M}$ a multi-label model with labels $\mathscr{L}= \{l_1,...,l_M\}$. $\mathscr{M}$ : $\mathcal{X} \subseteq \mathbb{R}^N \longrightarrow \mathcal{Y} \subseteq \mathbb{R}^M$. We have $N$ the input features size and $M$ the number of labels.
For each input example $x$, we denote by $\bar{y}$ the corresponding ground-truth and we have $\bar{y} = (y_1, ..., y_i, y_M)$ where $y_i \in \{0,1\}$ is the corresponding ground truth for label $i$. 

For each $x \in \mathcal{X}$, let $\hat{y}$ the predicted labels $\hat{y} = \mathscr{M}(x)$. 

Then, we denote by $acc_{k,\mathscr{M}}(x,\bar{y})$ the \textbf{k-accuracy of the input $x$} for its top $k$ labels, and define it as the cardinal of the intersection between $x$'s top-k ground truth labels and its top-k predicted labels:

$$  acc_{k,\mathscr{M}}(x,\bar{y}) = \frac{|(argsort_k (\hat{y})|) \cap (argsort_k (\bar{y})|)}{k}$$

where $argsort_k$ of a set are the indices of the top k elements of the set.
\end{definition}

For an input $x$, $acc_{k,\mathscr{M}}$ evaluates how much the most probable predicted labels match the most probable ground truth labels. This formalism is suitable for both ordinal labels (to take into account uncertainty) and multi-labels (to take into account label co-occurrence).

\begin{definition}

We define the \textbf{k-accuracy of the model $\mathscr{M}$} as the expectation over the input set $\mathcal{X}$ of the k-accuracy of the input $x \in \mathcal{X}$:
$
acc_{k,\mathscr{M}} = \mathbb{E}_{x}\left[ acc_{k,\mathscr{M}}(x,\bar{y}) \right]
$
\end{definition}

For k=1 the k-accuracy matches the standard accuracy.

\subsection{Experimental setup}


\paragraph{Datasets.} 

Following the protocol set up by \cite{cohen2020limits} we evaluate the robustness of CXR models using four datasets: 

\begin{itemize}
    \item NIH Chest X-ray14 \cite{Wang_2017}, denoted as \emph{NIH} in the following. A dataset of 112k images was labeled automatically with the NegBio labeler. This is the most common dataset used in the literature of CRX image classification.
    
    \item  \emph{CheXpert} \cite{irvin2019chexpert}. This dataset of 224k chest radiographs has been 
    labeled with a custom automated labeler over the NLP analysis of radiology reports. 
    
    \item \emph{PadChest} \cite{Bustos_2020} is a 160k image dataset. 
    The labels are extracted from radiographic reports manually annotated by trained physicians for 27\% of them. 
    
    \item A combination of the three denoted as \emph{AllD}. We combine the images obtained from the three previous datasets for this dataset and process them as proposed in \cite{cohen2020limits}. 
    
\end{itemize}

For each dataset, we evaluate the robustness using 5120 inputs randomly sampled from the test set. 

\paragraph{Models.}

All our models output a vector of 18 logits to cover the maximum number of labels of our evaluation, even if the dataset the model has been trained on is missing one or a few labels. This allows us to train and test each model on any other dataset. All our models have an average AUC $>$ 0.79. 

For the dataset specific models, we use pre-trained models using a DenseNet-121 architecture available in the TorchXrayvision library \cite{cohen2020limits}. It includes models trained on NIH, CheXpert (CHEX) and PadChest (PC). The library also provides pre-trained models on the MIMIC and RSNA datasets. Those are smaller CXR datasets that share the same labels as the \emph{AllD} dataset. 

We also compare the robustness of models with different architectures trained using the same dataset. We evaluate the performance of the DensetNet121 architecture and the Resnet512 architecture when trained using the \emph{AllD} dataset. 

Following similar work \cite{cohen2020limits, baltruschat2019comparison}, we adjust the training process to the CXR classification task: We account for the missing labels by training the models using only the loss from the available labels. CXR classification also suffers from a large imbalance in label distribution. We alleviate the imbalance with a frequency-based weight for each label: The less frequent labels have a higher contribution to the loss computation. Finally, each label also has a different optimal binary threshold. Except to evaluate the multi-label accuracy, we do not threshold the outputs and use the raw probabilities. For the multi-label accuracy, different thresholds are used for each label as proposed by Cohen et al.\cite{cohen2020limits}.

\paragraph{Attacks.}

We evaluate the robustness of the models mainly against PGD attack \cite{pgd_madry2019deep}. It has been shown by Madry et al. as a universal surrogate for first-order gradient attacks, and the robustness against PGD attacks is a common metric to evaluate the robustness of models \cite{croce2020robustbench}. It is also the one used in previous research about the robustness of CRX models \cite{advattackmedicalsystem, understandingadvattacks}.

We evaluate the 2 hyperparameters of PGD: The maximum perturbation size $\epsilon$ over the range of $\{0.5/255, 1/255, 2/255, 4/255, 8/255 \}$, and the number of attack steps in the range of $\{1, 5, 10, 25, 50\}$.



\paragraph{Robustness evaluation metrics.}

In addition to the k-robust accuracy, we also evaluate the robustness of the models using traditional error metrics, to cover metrics designed specifically for multi-label classification and ordinal classification: The mean square error (MSE), cross-entropy error (BCE), multi-label accuracy (MLACC) \cite{song2019multilabel} and the Ordinal classification loss (OL) \cite{frank2001simple}.

\section{Results and Evaluation}

\subsection{Cross-domain generalization}
\label{sec:rq1}

To better understand how adversarial attacks impact CXR classification models, we evaluate the impact of the training data on the robustness of models, in particular for transfer attacks, when the source model and the target models are different.
Given a PGD attack of $\epsilon=1/255$ and 25 steps, we evaluate the k-robust accuracy for k=1 and k=3 for our six DensetNet121 models M1, M2, M3, M4, M5, M6, and M7. The clean images are randomly sampled from the NIH dataset. 

\paragraph{Adversarial attacks transferability:}
Results are shown in table \ref{table:rq1-model-eval}.
When restricted to the 1-robust accuracy, the NIH model is the most robust model (15.4\% robust accuracy on average) and the CHEX model is the most vulnerable. When we evaluate the 3-robust accuracy, the most robust model becomes the AllD model (35.6\% robust accuracy on average). 
This confirms not only that different models have a large range of robustness (NIH is ten times more robust than CHEX), but previous claims that PGD attack on CXR classification yields a 100\% success rate are far from true. Taking into account the dataset used for model training can yield a significant difference in robustness.

The significant variability in the performances moving from top1 to top3 shows that actually, the models, in general, remain robust enough and the correct labels are still predicted with high probabilities. The only exception is the CHEX model, which remains very vulnerable. It hints that the distribution that has been learned by this model can significantly be impacted by a small perturbation.

Additionally, the most robust model in the white box threat model (the diagonal values where the source and target model are the same) is not the same given the 1-robust accuracy or the 3-robust accuracy. While the NIH model preserves the most probable class, the PC model retains better the correct labels in the top3 predictions. 


\begin{table}

\begin{center}
\scalebox{0.75}{
\begin{tabular}{|c|c|cccccc|}
\hline

Topk & Target $\rightarrow$& NIH & CHEX & PC & MIMIC & RSNA & AllD \\
Acc & Source $\downarrow$ &  &  &  &  &  &   \\
\hline
\multirow{5}{*}{k=1}& NIH &  \textbf{13.78} &  1.38 &  9.66 &  8.47 &   7.12 &  2.25\\
& CHEX &  15.66 &  1.62 &  9.28 &  8.00  &  7.09 &  2.44\\
& PC &  15.72 &  1.38 &  8.25 &    8.28 &  7.28 &  2.41\\
& MIMIC &  15.81 &  1.38 &  9.00 &  8.16 &  7.22 &  2.38\\
& RSNA &  15.78 &  1.31 &  9.97 &    8.06 &  8.09 &  2.53\\
& AllD &  15.72 &  1.34 &  9.34 &   8.09 &  7.28 &  4.78\\
\hline
\hline
\multirow{5}{*}{k=3}& NIH & 21.41 &  6.09 &  35.31 &   27.28 &  27.22 &   36.12\\
& CHEX & 21.66 &  6.28 &  34.88 &  27.47 &    26.91 &   36.28 \\
& PC & 21.69 &  6.41 &  \textbf{34.94} &   27.09 &  26.06 &   36.78\\
& MIMIC & 21.75 &  6.09 &  35.06 &   27.94 &  27.62 &   36.22\\
& RSNA & 21.59 &  6.12 &  35.41 &  27.00 &  21.16 &   36.25\\
& AllD & 21.69 &  6.34 &   35.38 &  27.12 &  27.34 &  32.12\\
\hline
\end{tabular}
}
\caption{k-robust accuracy on NIH dataset for six Densenet121 models, each trained on different chest x-ray datasets. The columns are the target models and the rows are the source models.
}
\label{table:rq1-model-eval}
\end{center}
\vspace{-4mm}
\end{table}

\paragraph{Robustness over different test datasets:}
Next, we explore the impact of the test dataset on the robustness of models. We evaluate in Table \ref{table:rq1-data-eval} the k-robust accuracy of each model when the original inputs are sampled from one of the datasets: D1 for  NIH Chest X-ray14, D2 for the CheXpert dataset, D3 for the PadChest dataset. The source and target models are the same in this setting.

Our results show that the examples sampled from the CheXpert dataset are the most vulnerable, except for the model trained on the NIH dataset. There is also no relationship between the robustness of the model and the distribution of inputs. Sampling the inputs for the adversarial examples from the same distribution (NIH with D1, CHEX with D2, PC with D3) does not reliably lead to higher robust accuracy.

Looking at the 3-robust accuracy, D3 is the more robust dataset for four models among the six. When the train examples and the evaluation example are both sampled from the PadChest dataset, the 3-robust accuracy peaks at 53.59\%, more than three times the robustness of the NIH model on the same dataset.  

\begin{table}

\begin{center}

\scalebox{0.85}{
\begin{tabular}{|c|ccc||ccc|}
\hline
k-robust acc  & \multicolumn{3}{c||}{k=1} & \multicolumn{3}{c|}{k=3} \\
\hline
  Dataset $\rightarrow$  & \multirow{2}{*}{D1} & \multirow{2}{*}{D2} & \multirow{2}{*}{D3} & \multirow{2}{*}{D1} & \multirow{2}{*}{D2} & \multirow{2}{*}{D3}  \\
  Model $\downarrow$ &  &  &  &  &  &   \\
\hline
NIH &  \underline{13.78} &  \underline{34.38} &   3.31 & 21.41 &  \underline{43.03} &  13.88 \\
CHEX & 1.62 & 0.88 &   1.31 & 6.28 & 9.03 &   19.47 \\
PC & 8.25 & 11.78 &    12.06 & \underline{34.94} & 22.66 &     \underline{53.59}\\
MIMIC & 8.16 & 5.72 &  \underline{17.47} & 27.94 & 40.84 &   49.25\\
RSNA & 8.09 & 5.38 &   7.22 & 21.16 & 17.00 &  16.38 \\
ALLD & 4.78 & 5.56 &    8.62 & 32.12 & 19.09 &  37.69 \\
\hline
\end{tabular}
}
\caption{k-robust accuracy on our three datasets for six Densenet121 models, each trained on different chest x-ray datasets. The source and target models for the attacks are the same. The columns are the datasets from which the example are sampled, and the rows are the source/target models. 
}
\label{table:rq1-data-eval}
\end{center}
\vspace{-4mm}
\end{table}

\paragraph{Impact of architecture:}

We observe in Table \ref{table:rq1-arch-eval} that different architectures are not reliably robust across different CXR datasets. While Resnet is more robust on D3, the DenseNet model has higher robust accuracy on D1. 
It is also noted that across both architectures, the dataset D3 yields the highest robust accuracy across both architectures. It is consistent with our previous results (Table \ref{table:rq1-data-eval}) that also showed that inputs from D3 are more robust across different models of the same architecture.

\begin{table}

\begin{center}
\begin{small}
\scalebox{0.85}{
{\renewcommand{\arraystretch}{1}
\begin{tabular}{|c|ccc||ccc|}
\hline
k-robust acc  & \multicolumn{3}{c||}{k=1} & \multicolumn{3}{c|}{k=3} \\
\hline
  Dataset $\rightarrow$  & \multirow{2}{*}{D1} & \multirow{2}{*}{D2} & \multirow{2}{*}{D3} & \multirow{2}{*}{D1} & \multirow{2}{*}{D2} & \multirow{2}{*}{D3}  \\
  Model $\downarrow$ &  &  &  &  &  &   \\
\hline
DenseNet121 & \underline{4.78}  &  \underline{5.56}  &  8.62 &  \underline{32.125}  & 19.09 &   37.69\\
Resnet50 & 3.31  & 2.56 &  \underline{11.44} & 22.25  & \underline{30.91} &   \underline{39.31}\\
\hline
\end{tabular}
}
}
\caption{k-robust accuracy on 2 architectures: Resnet50 and Densenet121. The source and target models for the attacks are the same. The columns are the datasets from which the example are sampled, and the rows are the source/target models. Greyed cells are best the values across a row, and  underlined cells are the best values across a column.
}
\label{table:rq1-arch-eval}
\end{small}
\end{center}
\vspace{-4mm}
\end{table}

\paragraph{Impact attack budget $\epsilon$ and $nb steps$:}

For k=3, robust accuracy drops from 34.56\% with $epsilon=0.5/255$ to 13\%  with $epsilon=4/255$. Meanwhile, the robust k-accuracy for k=1 slightly increases with the increased attack budget, from 3.84\% to 8.06\%. This increase in robustness is unexpected and can indicate that iteration budgets of 25 steps are not sufficient to effectively explore such a large search space.

Given a perturbation budget of $epsilon=0.5/255$, the number of steps has a limited impact on the robust accuracy. We observe that the attack success (and thus the models' robustness) plateaus around 30\% for all the multi-step attacks: 5, 10, 25, and 50 steps. 

\begin{mdframed}[style=MyFrame]
\textbf{Conclusion:} The robustness of CXR image classifiers significantly varies when considering architectures and datasets. Contrary to common practice, mixing multiple datasets leads to less robust models. 
\end{mdframed}

\subsection{Domain Specific knowledge}
\label{sec:rq2}

CXR classification not only raises questions about the generalization of one's hypothesis about the robustness of models, as we showed, but it also requires a higher understanding of the labels and diseases that we aim to classify. When dealing with a critical task like medical diagnosis, the risk associated with a prediction error can dramatically increase when the predicted diseases are far from the actual truth. We show that targeted adversarial attacks against these risky labels provide a new view of the robustness of CXR classification models.

To model the prediction risk, we use the co-occurrence matrix provided by the multi-label dataset NIH. In this dataset, each radiograph can have 1, 2 or 3 diseases that have been annotated. This matrix indicates which combination of diseases are very rare in practice, and hence can hardly be confused. For instance, while Infiltration and Atelectasis are two labels commonly found in annotations, Infiltration and Pneumothorax are scarce together. 






Let $\mathscr{M}$ a multi-label model with labels $\mathscr{L}= \{l_1,...,l_M\}$. $\mathscr{M}$ : $\mathcal{X} \subseteq \mathbb{R}^N \longrightarrow \mathcal{Y} \subseteq \mathbb{R}^M$.

For each $x \in \mathcal{X}$, let $\hat{y}$ the predicted labels $\hat{y} = \mathscr{M}(x) = (\hat{y}_1, \hat{y}_i..., \hat{y}_M) $ where $\hat{y}_i \in \mathbb{R}$ is the predicted probability of label $i$. Let $\hat{y}^*$ the most probable label for $x$: $\hat{y}^* =  \argmax_{i} \{\hat{y}_1, \hat{y}_i..., \hat{y}_M \}$. Let $C$ the normalized inverse co-occurrence matrix of the label space $\mathcal{Y}$. A higher value in $C$ means that the labels of the row and column indices are very unlikely to occur together. $C(i)$ is the vector of improbable labels associated with label $i$.

For each input $x \in \mathcal{X}$, we generate an adversarial $x^* \in \mathcal{X}$ example with targeted Projected Gradient Descent (PGD) \cite{pgd_madry2019deep} algorithm, targeted on the improbable label vector of $x$. Targeted PGD adds iteratively a perturbation $\delta$ that opposes the sign of the gradient $\nabla$ with respect to the input x and target $C(\hat{y}^*)$.  $\Pi$ is a clip function that ensures $x+\delta$ respects a $p$-$norm$ perturbation budget:
\begin{equation*}
x^0 = x \;;\;  x^{t+1} = \Pi_{x+\delta}(x^t - \alpha sgn(	\nabla_x L(\theta, x, C(\hat{y}^*))))
\label{eq:pgd}
\end{equation*}

This optimization can be seen as a weighted multi-label classification attack because the target vector is a real-valued vector, or as an ordinal classification attack because the order of the values of target logits actually matter. They reflect the risk caused by the misclassification. This risk can be computed as a vectorial product between the predicted logits of the adversarial example and the target logits computed with $C$. And we have: $RISK =  \mathscr{M}(x^*) \times C(\hat{y}^*) $.

To account for both views, we use different loss functions: MSE, BCE, OL. We report for each approach the MSE, BCE, AUC, MLACC, and RISK. We also include the k-robust accuracy (with k=1,3) to compare this threat model with the threat model of \ref{sec:rq1}. While MSE, BCE, AUC, and k-robust accuracy reflect how much error we introduce in comparison with the original prediction, MLACC and RISK reflect how close is the predicted output to the target output.
We bring together the results of all these evaluations in Table \ref{table:rq2-model-eval}.

\paragraph{Impact of the loss function}

The most robust models are overall consistent across different loss functions used in the attack. This confirms that handling risk-based attacks as an ordinal classification problem are as relevant as a multi-label problem for the success of the attack    

\paragraph{Impact of the threat model}

Comparing the k-robust accuracy of our risk-based threat model with the untargeted threat model of our previous results (Table \ref{table:rq1-model-eval}) shows that this risk-based threat model yields more successful attacks, and thus lower robust accuracy of the models. 

For instance, with k=1, the robust accuracy of the NIH model against untargeted attacks is 13.78\% (Table \ref{table:rq1-model-eval}), but it drops to 2.69\% under the risk-based attacks (Table \ref{table:rq2-model-eval}). Similarly, for k=3, the AllD model has a robust accuracy of 32.12\% against untargeted attacks and only 11.12\% against risk-based attacks.

\paragraph{Risk evaluation of the models}

According to the RISK metric, the NIH model is not only the most robust to adversarial attack but also the one where the end labels have the lowest probabilities to actually be rare co-occurring labels of the original label.

\begin{table}

\begin{center}
\scalebox{0.78}{
\begin{tabular}{|c|c|cccccc|}
\hline

 & Model $\rightarrow$& NIH & CHEX & PC & MIMIC & RSNA & AllD \\
Loss & Metric &  &  &  &  &  &   \\
\hline
\multirow{7}{*}{MSE}& k=1 $\uparrow$ & 3.07 & 0.25  & 0.19  & 0.25  &  0  & \cellcolor{lightgray} 4.19\\
& k=3 $\uparrow$&  \cellcolor{lightgray}13.31 & 3.31  & 16.13  & 2.63  &  0  &  10.82 \\
& AUC $\uparrow$ & 0.74 & 0.5  & \cellcolor{lightgray}0.78  & 0.5  & 0.5    & 0.42\\
& MSE $\downarrow$ & 0.07 & 0.06  &  0.13 & 0.09 &  \cellcolor{lightgray}0.02 &    0.11  \\
& BCE $\downarrow$ &  0.78 & 0.77  & 0.85  & 0.76  &  \cellcolor{lightgray} 0.74  & 0.84\\
\cline{2-8}
& MLACC $\downarrow$ & 0.66 & 0.58  & 0.71 & \cellcolor{lightgray}0.43  & 0.45   & 0.59 \\
& RISK $\downarrow$ & \cellcolor{lightgray}0.18 & 0.2  & 0.28  &  0.25 &  0.24  & 0.23\\
\hline
\hline
\multirow{7}{*}{BCE}& k=1 $\uparrow$& 2.69  & 1.19  &  0.19 &  0.69 &  0  &  \cellcolor{lightgray}3\\
& k=3 $\uparrow$&  \cellcolor{lightgray}12.43& 8.19  & 16.5  & 1.75  &   0 & 11.12\\
& AUC $\uparrow$& \cellcolor{lightgray}0.78 & 0.5  & 0.81  & 0.5  & 0.5    & 0.52 \\
& MSE $\downarrow$& 0.07  & 0.06  &  0.12 &  0.1 &  \cellcolor{lightgray}0.01  & 0.09 \\
& BCE $\downarrow$& 0.77 & 0.76  & 0.84  & 0.76  &  \cellcolor{lightgray}0.74  & 0.81 \\

\cline{2-8}
& MLACC $\downarrow$& 0.67 & 0.59  & 0.71  &  \cellcolor{lightgray}0.43 & 0.45 & 0.6 \\
& RISK $\downarrow$&  \cellcolor{lightgray}0.17 & 0.19  &  0.27 &  0.24 & 0.24 & 0.26 \\
\hline
\hline
\multirow{7}{*}{OL}& k=1 $\uparrow$& \cellcolor{lightgray}2.81 & 0.31  & 0.63  & 0.13  & 0   & 2.5\\
& k=3 $\uparrow$&  \cellcolor{lightgray} 14.06 & 5.31  &  14.56 & 1.06  &  0  & 9.88\\
& AUC $\uparrow$&  \cellcolor{lightgray}0.78 & 0.5  & 0.69  &  0.5  &  0.5  & 0.41 \\
& MSE $\downarrow$& 0.07 & 0.08  & 0.12  & 0.09  &  \cellcolor{lightgray}0.02  & 0.11\\
& BCE $\downarrow$& 0.78 & 0.78  & 0.84  & 0.76  &  \cellcolor{lightgray}0.74  & 0.84\\
\cline{2-8}
& MLACC $\downarrow$& 0.66 & 0.53  & 0.68  &  \cellcolor{lightgray}0.42 &    0.45& 0.54 \\
& RISK $\downarrow$& \cellcolor{lightgray}0.17 &  0.2 &  0.27 & 0.24  & 0.24   & 0.22 \\
\hline

\end{tabular}
}

\caption{Robustness of metrics for six Densenet121 models on NIH dataset, attacked using three loss functions. $\downarrow$ and $\uparrow$ indicate that lower (higher respectively) is more robust.
}
\label{table:rq2-model-eval}
\end{center}
\vspace{-4mm}
\end{table}

\paragraph{Impact of the robustness metric}

Our results show that the error metrics fail to highlight one specific model as being the most robust. According to robust accuracy and robust AUC, NIH is the most robust model across different loss functions. Meanwhile, the RSNA model is the most robust according to the BCE and MSE losses.





We also evaluate the Pearson correlation between the robustness values of each batch of all models combined. 
Except for the correlation between the risk and the 3-robust accuracy, the p-value is under 10\^-3 
. Our results show that none of the existing metrics (MSE, BCE,...) is correlated with the RISK metric. This confirms that existing metrics do not take into account this dimension. As expected, MSE and BCE are highly correlated with each other and mildly correlated with the top-3 robust accuracy. On the contrary, top-1 robust accuracy has little correlation with the other metrics. 

\begin{mdframed}[style=MyFrame]
\textbf{Conclusion:} The choice of a CXR classification model can significantly vary based on the robustness metric and threat models we evaluate. 
\end{mdframed}



\section*{Conclusion}

Evaluating the robustness of chest radiograph classifiers requires extreme caution and consideration when designing the experimental protocol.
Our study, in particular, outlines the following recommendations. State a precise and realistic threat model for your use case. Use clear assumptions about the distribution learned by the target model and how it relates to what your craft model has learned. Choose the right robustness metric depending on the task and threat model: Single-label classification, multi-label classification, and ordinal classification metrics reflect different vulnerabilities. Identify the risks and their impacts, and evaluate the robustness of robust models use strong baselines and risky attack strategies, e.g., with larger attack budgets.

We do not intend for this paper to be the definitive answer, nor that the items contained above are exhaustive
. We encourage future research to confront their protocol with actual use cases, and we hope that our work paves the way to critical thinking about the protocols of adversarial attack evaluation in the real world.

{\footnotesize
\bibliography{medicalbib,advbib}}


\end{document}